\documentclass[conference]{IEEEtran}
\IEEEoverridecommandlockouts
\usepackage{multirow}
\usepackage{threeparttable}
\usepackage{algorithmic}
\usepackage{graphicx}
\usepackage{textcomp}
\usepackage{xcolor}
\usepackage{stfloats}
\usepackage{bbding}
\usepackage{pifont}
\usepackage{hyperref}
\usepackage{makecell}
\usepackage{booktabs}
\usepackage{tabularx}
\usepackage{amsmath}
\usepackage{array}
\usepackage{subcaption}
\usepackage[most]{tcolorbox}
\usepackage{times}
\usepackage{latexsym}
\usepackage[T1]{fontenc}
\usepackage[utf8]{inputenc}
\usepackage{microtype}
\usepackage{inconsolata}
\usepackage{colortbl}
\usepackage[framemethod=TikZ]{mdframed}
\definecolor{mycolor}{RGB}{194, 214, 236}
\usepackage{cite}
\usepackage{amsmath,amssymb,amsfonts}

\newcounter{finding}
\newcommand{\finding}[1]{\refstepcounter{finding}
 	\vspace{1mm}
	\begin{mdframed}[linecolor=gray,roundcorner=12pt,backgroundcolor=gray!15,linewidth=3pt,innerleftmargin=10pt,innertopmargin=6pt,innerbottommargin=6pt,leftmargin=0cm,rightmargin=0cm,topline=false,bottomline=false,rightline = false]
		\textbf{Findings \arabic{finding}:} #1
	\end{mdframed}
	\vspace{0.5mm}
}

\newcommand{\cxmark}{\ding{51}\textsuperscript{\kern-0.55em\ding{55}}}
\definecolor{dgreen}{RGB}{46,175,87}

\makeatletter 
\newcommand{\thickhline}{%
    \noalign {\ifnum 0=`}\fi \hrule height 1pt
    \futurelet \reserved@a \@xhline
}
\makeatother

\def\BibTeX{{\rm B\kern-.05em{\sc i\kern-.025em b}\kern-.08em
    T\kern-.1667em\lower.7ex\hbox{E}\kern-.125emX}}

\begin{document}

\title{How Far Have We Gone in Binary Code Understanding Using Large Language Models}

\newcommand\corrauthorfootnote[1]{%
  \begingroup
  \renewcommand\thefootnote{}\footnote{\textsuperscript{\dag}#1}%
  \addtocounter{footnote}{-1}%
  \endgroup
}

\author{Xiuwei Shang\textsuperscript{1},
Shaoyin Cheng\textsuperscript{1,2,\dag},
Guoqiang Chen\textsuperscript{1},
Yanming Zhang\textsuperscript{1},
Li Hu\textsuperscript{1},
Xiao Yu\textsuperscript{1}, 
Gangyang Li\textsuperscript{1}, \\
Weiming Zhang\textsuperscript{1,2},
Nenghai Yu\textsuperscript{1,2}  \\
\texttt{\{shangxw,ch3nye,azesinter,pdxbshx,yuxiao1217,ligangyang\}@mail.ustc.edu.cn} 
\\
\texttt{\{sycheng,zhangwm,ynh\}@ustc.edu.cn}
\\
\textsuperscript{1}University of Science and Technology of China, Hefei, China\\
\textsuperscript{2}Anhui Province Key Laboratory of Digital Security, Hefei, China  \\
}

\maketitle

\begin{abstract}
Binary code analysis plays a pivotal role in various software security applications, such as software maintenance, malware detection, software vulnerability discovery, patch analysis, etc. However, unlike source code, understanding binary code is challenging for reverse engineers due to the absence of semantic information. Therefore, automated tools are needed to assist human players in interpreting binary code. 
In recent years, two groups of technologies have shown promising prospects: (1) Deep learning-based technologies have demonstrated competitive results in tasks related to binary code understanding, furthermore, (2) Large Language Models (LLMs) have been extensively pre-trained at the source-code level for tasks such as code understanding and generation. This makes participants wonder about the ability of LLMs in binary code understanding.

In this work, we propose a benchmark to evaluate the effectiveness of LLMs in real-world reverse engineering scenarios. The benchmark covers two key binary code understanding tasks, including function name recovery and binary code summarization. We gain valuable insights into their capabilities and limitations through extensive evaluations of popular LLMs using our benchmark. Our evaluations reveal that existing LLMs can understand binary code to a certain extent, thereby improving the efficiency of binary code analysis. Our results highlight the great potential of the LLMs in advancing the field of binary code understanding. 

\end{abstract}


\section{Introduction}
\corrauthorfootnote{Corresponding Author.}
Binary code analysis is fundamental to software security, serving as the bedrock technology for many critical tasks including reverse engineering \cite{Canfora2011}, software vulnerability detection \cite{Giffin2004EfficientCI}, and malware analysis \cite{Alrabaee2022}. The process of compilation, however, leads to the elimination of semantic information present at the source-code level. Additionally, binary files often have their symbol information stripped \cite{Zhang2021sp} for various reasons (e.g., copyright protection \cite{Meng2016issta} and obscuring functionality \cite{Patrick2020acsac}). Thus, it is challenging for reverse engineers to understand the semantics of binary code. 

Although many decompilation tools, such as IDA Pro \cite{IDA}, Ghidra \cite{Ghidra} and BinaryNinja \cite{BinaryNinja}, can heuristically convert binary code into C-like pseudo code, they still lack easy-to-understand semantics information, especially function names and code comments that play an important role in comprehending the code \cite{Gellenbeck1991, nfre2021issta}. 
Recently, borrowing ideas from Natural Language Processing (NLP), deep learning-based methods have been proposed for understanding binary code. In the task of function name recovery, NERO \cite{nero2020}, NFRE \cite{nfre2021issta} and SymLM \cite{symlm2022ccs} utilized the disassembled assembly instruction sequence as neural models input to reassign descriptive names. NER \cite{Chen2023pst} utilized decompiled pseudo code with a higher abstraction level as input and achieves better performance.

Besides, the function name alone is not enough to represent the complete behavior of the code \cite{Sridhara2010ase}. If a natural language description can be generated for the binary code, it will greatly save the reverse engineer's analysis time. BinT5 \cite{bint52023saner} is the first generation model designed for binary code summarization, which is based on the source code model CodeT5 \cite{wang-etal-2021-codet5} and fine-tuned on binaries. As a unified pre-training model, HexT5 \cite{Xiong2023ase} can perform multiple downstream tasks such as code summarization and function name recovery. However, the methods mentioned above generally exhibit limited generalization to unseen binary code. 

Recently, Large Language Models (LLMs) have attracted significant attention in the academic community. 
General LLMs, such as Llama \cite{Llama2023llama}, ChatGPT \cite{chatgpt2022training}, etc., have been widely demonstrated for their capabilities in natural language understanding and generation. Furthermore, LLMs in the programming language domain (e.g., CodeLlama \cite{codellama2023rozière} and WizardCoder \cite{wizardcoder2023luo}) have shown notable ability in program analysis tasks, like fixing security vulnerabilities \cite{HowEffective_ISSTA_2023}, test cases auto-generation \cite{zhang2023sectests}. These developments demonstrate the potential of LLMs to handle complex and structured information flows that are particularly important for understanding binary code. 

In this paper, instead of developing a new technique, we investigate and compare the capabilities of various LLMs in understanding binary code. By harnessing the advanced analytical power of LLMs, we seek to explore the extent to which these models are able to understand binary code, a task that traditionally handled by skilled human engineers \cite{David2020Neural}. To facilitate our evaluation, we designed an automated approach to construct an evaluation benchmark dataset, which includes aligned source code, natural language summaries, and decompiled pseudo code. We then contrasted the capabilities of LLMs on two binary code understanding tasks, namely: (1) function name recovery, and (2) binary code summarization. We extensively evaluated eight code domain LLMs (CodeGen \cite{codegen2_2023nijkamp}, WizardCoder \cite{wizardcoder2023luo}, DeepSeek-Coder \cite{deepseekcoder}, CodeLlama \cite{codellama2023rozière} et.al.), eight general domain LLMs (ChatGLM \cite{chatglm2023glm130b}, Vicuna \cite{vicuna2023judging}, Llama \cite{Llama2023llama}, Mistral \cite{mixtral2024jiang}, ChatGPT \cite{chatgpt2022training} et.al.), and four deep learning-based expert models (SymLM \cite{symlm2022ccs}, NER \cite{Chen2023pst}, BinT5 \cite{bint52023saner}, HexT5 \cite{Xiong2023ase}). Additionally, we explored the impact of injecting domain knowledge by fine-tuning LLMs on specific tasks. Furthermore, we conducted case studies in the context of virus analysis to showcase the performance of the LLMs in understanding binary code in real-world scenarios.

Our findings demonstrate that LLMs exhibit excellent potential in advancing automated binary code understanding. We call for more research in this area to further enhance the capabilities of LLMs to play a more critical role in the complex task of binary code analysis.

Our contributions can be summarized as follows:

\begin{itemize}
    \item We design an automated method to construct a benchmark dataset to evaluate the capabilities of binary code understanding and release it to facilitate further research.
    \item We conduct a thorough study that evaluates the capabilities of eight code domain LLMs, eight general domain LLMs, and four DL-based methods on binary code understanding. Our primary focus lies on two fundamental tasks: function name recovery and binary code summarization.
    \item Our findings provide valuable insights into the capabilities and limitations of LLMs for understanding binary code. We thoroughly discuss the outcomes of our evaluations and offer suggestions for future research directions, aiming to propel advancements in this domain.
\end{itemize}

\section{Background and Motivation \label{sec:background}}

\begin{figure*}[t]
	\centering
        \scalebox{0.90}{
	\includegraphics[width=\linewidth]{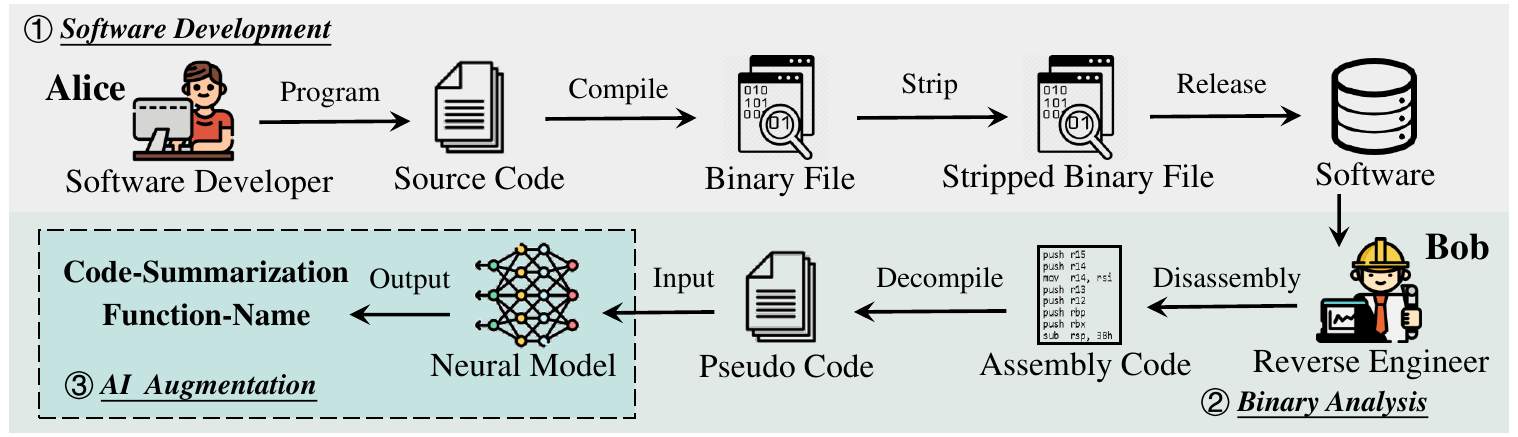}
        }
        \vspace{0ex}
	\caption{Application background of binary code understanding.}
    \vspace{-1ex}
    \label{fig:background}
\end{figure*}

\subsection{Binary Code Understanding}

\begin{figure}[t]
	\centering
        \scalebox{1}{
	\includegraphics[width=\linewidth]{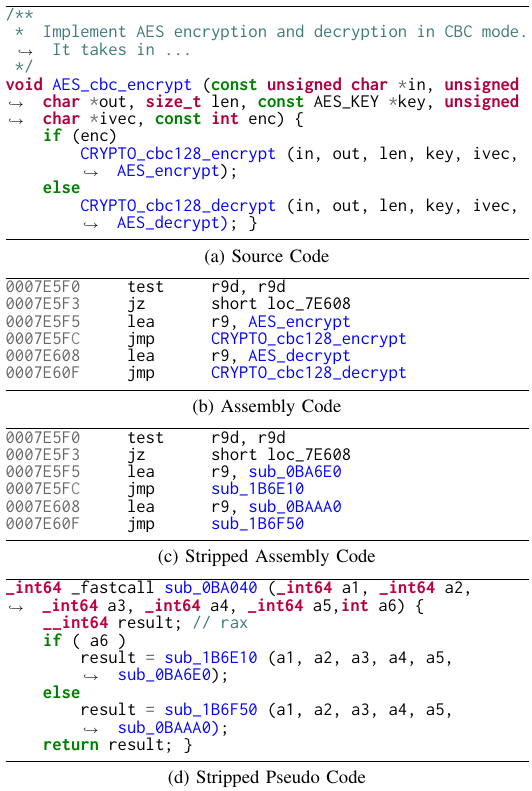}
        }
        \vspace{0ex}
	\caption{Example of source, assembly, stripped assembly and stripped pseudo code snippet.}
    \vspace{-3ex}
    \label{fig:code}
\end{figure}

As shown in Figure \ref{fig:background}, consider a scenario where a software developer, Alice, who writes a program in a high-level language like C/C++. Her code, written in a human-readable format (Figure \ref{fig:code}a), must be translated into a form the computer can execute. This is where the compilation process comes into play, turning her source code into binary code, a series of 1s and 0s that the machine understands.

Once compiled, Bob, a reverse engineer, wants to understand how Alice's program works. He uses a disassembler to convert the binary code back into an assembly code sequence (Figure \ref{fig:code}b), which is more readable than binary but still quite low-level. However, Alice has used the “strip” command to remove symbol information from the original binary code in order to reduce file size and protect intellectual property. This makes Bob's job more difficult because he is now missing critical information such as function names (replaced with meaningless placeholders such as \texttt{sub\_0BA6E0}) and code comments (Figure \ref{fig:code}c).

Finally, Bob uses a decompiler in an attempt to reconstruct the high-level logic of the program. The decompiler generates pseudo code (Figure \ref{fig:code}d), an approximation of what Alice's original source code might have looked like. However, due to the complexity of the decompilation process and the lack of symbolic information in the stripped binary, the pseudo code may not exactly match Alice's original code, making it still difficult for Bob to understand the function of the program.

At this point, Bob attempts to use advanced natural language processing (NLP) techniques, such as LLMs or deep learning models, which are adept at identifying patterns and inferring logical structures. Bob leverages these techniques to \textbf{predict function names} and \textbf{generate natural language summaries} of the code's functionality. This process can be formalized as:
\begin{equation}
\small
\mathcal{N},\ \mathcal{S}=f(\mathcal{F}, \mathcal{B}) 
\end{equation}
where $\mathcal{F}$ is a stripped decompiled function in pseudo code form in binary file $\mathcal{B}$, which is fed into LLMs denoted as $f$. The objective is to generate a meaningful function name denoted as $\mathcal{N}$, and a natural language description denoted as $\mathcal{S}$ of this function.

Through this process, Bob combined the analytical power of AI with his reverse engineering skills to bridge the gaps left by the stripped binaries and gain a deeper understanding of Alice's original programming intent.

\subsection{Related Works}

\subsubsection{Function Name Recovery}
This task aims to predict and recover descriptive function names for functions in binary files without symbolic information, which is of great significance for summarizing function semantics and understanding software behavior. Several methods based on deep learning have been proposed. Among them, NERO \cite{nero2020} uses augmented control flow graphs, combined with the neural model of the encoder-decoder paradigm, to effectively capture the behavioral characteristics of functions and provide a new method for recovering binary function names. NFRE \cite{nfre2021issta} proposes a lightweight framework for function name recovery that utilizes the sequential and structural information of assembly instructions. The efficiency and scalability of the framework provide the possibility to process a large-scale binary files. Based on NFRE, SymLM \cite{symlm2022ccs} further considers the calling context to help the model understand function semantics, and leverages advanced pre-training models \cite{pei2021trex} for instruction embedding. Finally, NER \cite{Chen2023pst} started from the perspective of binary code representation and studied the effectiveness of different representations for function name recovery using deep neural models, providing new perspectives and tools for this field.

\subsubsection{Binary Code Summarization}
Code summarization aims to extract key information from binary code and generate concise summaries, which is important for supporting program understanding. BinT5 \cite{bint52023saner} is the first model focused on binary code summarization, which extends the application scope of source code pre-trained language models. This model treats the decompiled code as a special programming language, uses fine-tuned CodeT5 \cite{wang-etal-2021-codet5} to capture its semantics and generate a summary. The introduction of BinT5 opens up new avenues for binary code summarization research. Furthermore, HexT5 \cite{Xiong2023ase} proposed a unified pre-training model also based on CodeT5, which allowed multi-task learning, supported function name recovery, binary code summarization and other downstream tasks, and demonstrated promising performance.

\subsection{Our Motivation}
Large Language Models (LLMs), often composed of billions or even trillions of parameters, are built upon large amounts of data to learn the relationship between program language and natural language. 
Recently, A few studies \cite{HowEffective_ISSTA_2023, zhang2023sectests, hou2023large} find that LLMs have demonstrated excellent capabilities in dealing with natural language tasks, as well as source code understanding, indicating that they have the potential to be applied to complex analysis of source code. 
Given this, we speculate that these models may be equally applicable to binary code understanding. Since binary code is similar in nature to source code and natural language, they all follow certain patterns and structures that can be learned and exploited by LLMs \cite{Zhang2023lev}.

Therefore, this study will explore the potential of LLMs in understanding binary code, aiming to evaluate whether these models can cross domain boundaries and extend their capabilities in natural language and source code to binary code analysis. This is expected not only to provide new perspectives for automated code understanding, but also to open up new application paths in areas such as reverse engineering and malware analysis.

\section{Evaluation Design}

\subsection{Dataset Construction}

\begin{figure*}[t]
	\centering
        \scalebox{0.93}{
	\includegraphics[width=\linewidth]{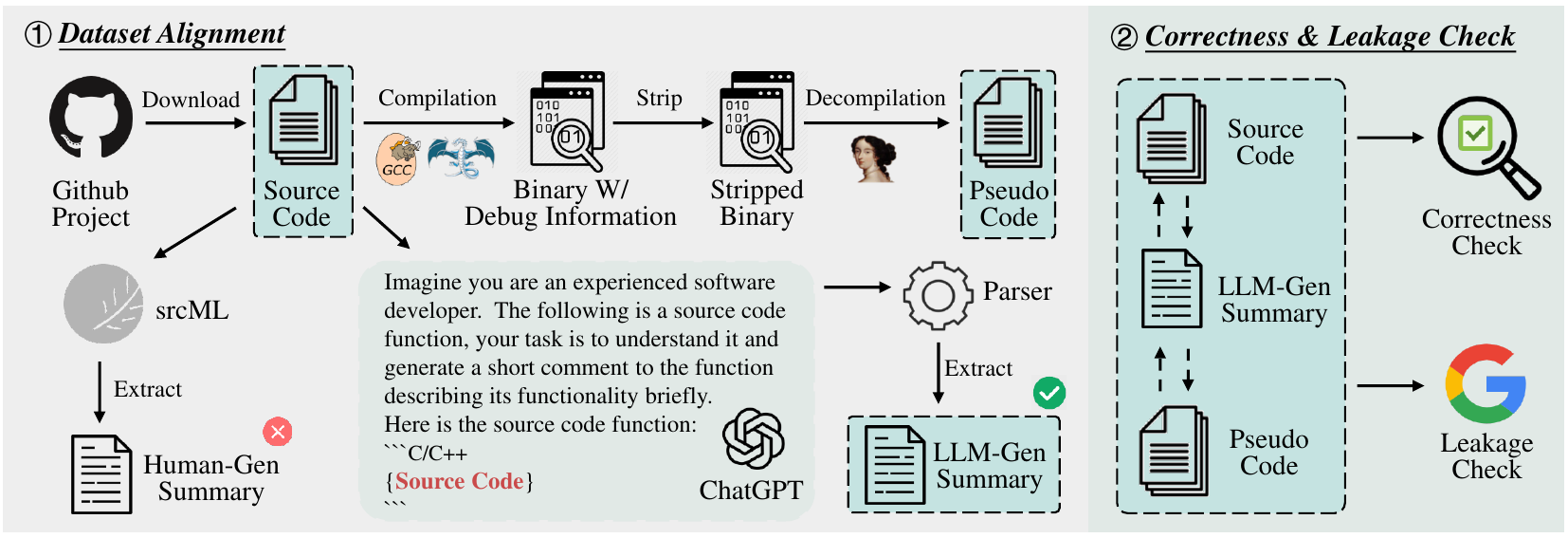}
         }
	\caption{An overview of the benchmark dataset construction process.}
    \vspace{-1ex}
    \label{fig:datacon}
\end{figure*}

\begin{table}
    \centering
    \setlength{\tabcolsep}{0.95mm}
    \scalebox{0.98}{
    \begin{tabular}{lcccc}
        \toprule
            \textbf{Project} & \textbf{Domain} & \textbf{\# Binaries} & \textbf{\# Functions} & \textbf{\# Select}\\
        \midrule
            FFmpeg \cite{FFmpeg} & Video  & 2 & 30,585 & 200 \\
            Redis \cite{Redis} & Database & 2 & 6,837 & 200 \\
            Curl \cite{Curl} & Network & 2 & 6,688 & 200 \\
            Masscan \cite{Masscan} & Network & 1 & 536 & 150 \\
            Llama2.c \cite{Llama2.c} & Neural Network & 2 & 35 & 34 \\
            Whisper.cpp \cite{whisper.cpp} & Neural Network & 9 & 5,339 & 200 \\
            OpenSSL \cite{OpenSSL} & Crypto & 2 & 19,122 & 300 \\
            zstd \cite{zstd} & Compress & 1 & 2,432 & 200 \\
            ImageMagick \cite{ImageMagick} & Image & 3 & 3,798 & 200 \\
            Libvips \cite{Libvips} & Image & 1 & 5,424 & 200 \\
            Libexpat \cite{Libexpat} & Format & 1 & 362 & 100 \\
            Ultrajson \cite{Ultrajson} & Format & 2 & 27 & 16 \\
        \midrule
            \textbf{Total (12)} & 8 & 28 & 81,185 & 2,000 \\
        \bottomrule
    \end{tabular} }
    \vspace{0ex}
    \caption{Statistics of our benchmark dataset.}
    \label{tab:dataset_construction} 
\end{table}

Before we can effectively evaluate the ability of LLMs to understand binary code, a comprehensive benchmark dataset is necessary to provide a consistent basis for different model evaluations and comparisons. The specific process of constructing the benchmark dataset is shown in Figure \ref{fig:datacon}. 

\subsubsection{\textbf{Source code selection}} To reflect real-world reverse engineering needs, we believe that code sources of the benchmark should meet the following requirements:

\begin{itemize}
	
	\item \emph{Authenticity}: the code should come from real projects, not toy programs or incomplete code snippets. 
	
	\item \emph{Breadth}: the selected code should cover multiple fields and not be limited to specific fields or application scenarios. 
	
	\item \emph{High quality}: the selected code should be of good quality, including clear structure, reasonable naming conventions, etc. 

    \item \emph{Credibility}: the selected code should ideally be sourced from projects maintained by well-known or reputable developers or organizations to accurately reflect real-world application scenarios.
\end{itemize}

Therefore, as shown in Table~\ref{tab:dataset_construction}, we select 12 real-world projects implemented in C language with the highest star ratings on Github, which have high credibility, excellent code quality and maintenance standards, covering eight application domains, including crypto, compress, network, video, image, database, neural network, etc.

\subsubsection{\textbf{Compile, strip and decompile}} We compile these projects on a machine running Ubuntu 22.04 OS, targeting the \textit{x86\_64} architecture, and for compiler and optimization level we use the default configuration of each project. As illustrated in Table~\ref{tab:dataset_construction}, we generate a total of 28 binaries. Subsequently, we employ the \texttt{strip} command in Linux to remove the symbol tables from these binaries. 

Previous research \cite{Chen2023pst} has found that using pseudo code as a representation of binary code is more effective for neural models than assembly instruction sequence and Intermediate Representation (IR). Therefore, we directly utilize IDA Pro\cite{IDA} to decompile the binary files and convert the binary code into pseudo code form without considering other representation forms.

\subsubsection{\textbf{Alignment}}  We use DWARF \cite{Dwarfformat} debugging information to align source code and pseudo code, which can record functions, variables in binary functions, and their locations (include source file name, line number, and column number) in source code. As shown in Table~\ref{tab:dataset_construction}, we obtain a total of 81,185 functions matching source code and pseudo code. To align source code and human-written summary, we use srcML \cite{Maletic2015Exploration} to analyze and parse the source files, then collect single- and multi-line summaries above the location of function declarations and definitions. This completes the alignment of source code -- pseudo code -- human summary.

\subsubsection{\textbf{Ground-truth identifiction}} For the function name recovery task, we parse the function names in the source code as labels using regular expressions. For the binary code summarization task, we first consider using human-written comments extracted from source code files as labels as in previous work \cite{bint52023saner, Xiong2023ase}. However, we found that only about 14.8\% of functions have comments written by human developers. Worse yet, not all comments are describing the functional summary of the function, but will also contain some noisy content, and they are of varying quality and style. Therefore, using human-written comments as ground-truth is unreliable.

Presently, an increasing number of research works \cite{Dagdelen2024, Bzdok2024, tan2024large} employ large language models such as ChatGPT \cite{chatgpt2022training} for tasks like data annotation, and has demonstrated a certain degree of reliability. Inspired by these pioneering works, we utilize ChatGPT to generate summaries as ground-truth. Specifically, we use the source code of the function to construct the prompt shown in Figure \ref{fig:datacon}, prompting ChatGPT to generate a short summary describing the function's purpose and functionality.

\subsubsection{\textbf{Correctness \& leakage check}}
It is crucial to ensure the correctness of the ground-truth, so we perform a correctness check on the descriptive summaries generated by ChatGPT. Specifically, we invited three experienced domain experts to review the match between the source code and the summary. Experts were asked to give each abstract a "pass" or "fail" score. If two or more experts give a "fail" rating, the data will be removed directly from the dataset; if one expert gives a "fail" rating, we will conduct a collective discussion and give a final in conclusion. We finally select 2,000 pieces of data as the benchmark dataset.

It is also imperative that benchmark datasets are not included in the training set of LLMs to mitigate the risk of data leakage. All our evaluation data are decompiled pseudo code, and the symbolic information is stripped away so that it is significantly different from the corresponding source code form, which greatly avoids data leakage. To further ensure the validity and reliability of our benchmark evaluation, we use the Google search engine to check whether the code appears on the Internet in clear text. The results showed that none of the pseudo codes were retrieved by whole-word matching. 

\subsection{Large Language Models Setup}

\begin{figure*}[t]
	\centering
        \scalebox{0.95}{
	\includegraphics[width=\linewidth]{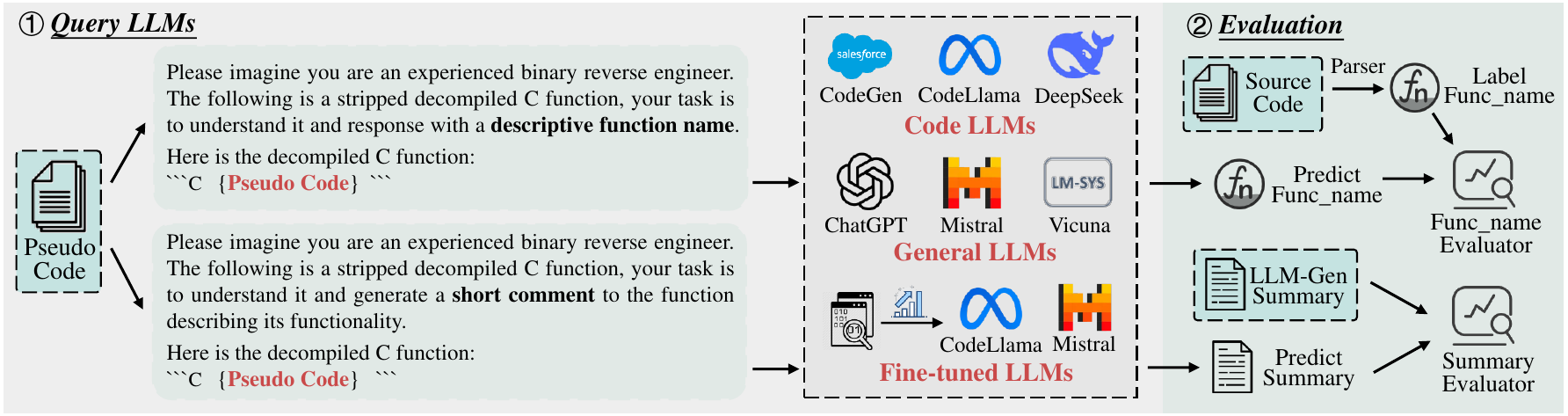}
         }
	\caption{An overview of the evaluation process.}
    \label{fig:eva_overview}
\end{figure*}

\begin{table*}[t]
  \centering
  \renewcommand{\arraystretch}{1.1}
  \setlength{\tabcolsep}{0.7mm}{
  \scalebox{0.83}{
    \begin{tabular}{clcccccccc}
    \toprule
        \multirow{2}{*}{\textbf{Domain}} & \multicolumn{1}{c}{\multirow{2}{*}{\textbf{Model}}} & \multirow{2}{*}{\textbf{\makecell{Release\\Time}}}  & \multirow{2}{*}{\textbf{\makecell{Size}}} & \multirow{2}{*}{\textbf{\makecell{Base\\Model}}} & \multicolumn{3}{c}{\textbf{Training Corpus}} & \multirow{2}{*}{\textbf{Publisher}} & \multirow{2}{*}{\textbf{License}}  \\
        \cmidrule(r){6-8}
        & & & & & \textbf{Raw Size} & \textbf{\#Tokens} & \textbf{\#Instances} & \\
    \midrule
     \multicolumn{1}{c}{\multirow{8}{*}{\makecell{Code \\ LLMs}}} 
       & CodeGen25-7b-instruct \cite{codegen2_2023nijkamp} & Jul-2023 & 7B & CodeGen2 & - & 1.4T & - & Salesforce & Open-source  \\
       & WizardCoder-15b-V1.0 \cite{wizardcoder2023luo} & Jun-2023 & 15B & StarCoder & - & - & 78.0K & WizardLM & Open-source \\
       & WizardCoder-33b-V1.1 \cite{wizardcoder2023luo} & Jan-2024 & 33B & Deepseek-Coder & - & - & - & WizardLM & Open-source \\
       & Code Llama-7b-instruct-hf \cite{codellama2023rozière} & Jun-2023 & 7B & Llama-2-7b & 4.4TB & 525.0B & - & Meta AI & Open-source \\
       & Code Llama-13b-instruct-hf \cite{codellama2023rozière} & Jun-2023 & 13B & Llama-2-13b & 4.4TB & 525.0B & - & Meta AI & Open-source  \\
       & Code Llama-34b-instruct-hf \cite{codellama2023rozière} & Jun-2023 & 34B & Llama-2-34b & 4.4TB & 525.0B & - & Meta AI & Open-source  \\
       & Code Llama-70b-instruct-hf \cite{codellama2023rozière} & Jan-2024 & 70B & Llama-2-70b & - & 1.0T & - & Meta AI & Open-source  \\
       & DeepSeek-Coder-33b-instruct \cite{deepseekcoder} & Nov-2023 & 33B & - & - & 2.0T & -  & DeepSeek-AI & Open-source  \\
    \midrule
      \multicolumn{1}{c}{\multirow{7}{*}{\makecell{General \\ LLMs}}} 
       & ChatGLM2-6B \cite{chatglm2023glm130b} & Jun-2023 & 6B & - & - & 1.4T & - & THUDM & Open-source \\
       & Vicuna-7b-v1.5 \cite{vicuna2023judging} & Aug-2023 & 7B & Llama-2-7b & - & - & 125.0K & L.Zheng et al. & Open-source \\
       & Vicuna-13b-v1.5 \cite{vicuna2023judging} & Aug-2023 & 13B & Llama-2-13b & - & - & 125.0K & L.Zheng et al. & Open-source \\
       & Llama-2-13b-chat-hf \cite{Llama2023llama} & Jul-2023 & 13B & - & - & 2.0T & - & Meta AI & Open-source \\
       & Llama-2-70b-chat-hf \cite{Llama2023llama} & Jul-2023 & 70B & - & - & 2.0T & - & Meta AI & Open-source  \\
       & Mistral-7B-Instruct-v0.2 \cite{mixtral2024jiang} & Dec-2023 & 7B & Mistral-7B & - & - & - & Mistral AI & Open-source  \\
       & Mixtral-8x7B-Instruct-v0.1 \cite{mixtral2024jiang} & Dec-2023 & 47B & Mistral-7B & - & - & - & Mistral AI & Open-source  \\
       & ChatGPT \cite{chatgpt2022training} & Nov-2022 & - & - & - & - & - & OpenAI & Closed-source \\
    \bottomrule
    \end{tabular} } }
    \caption{Detail information of Large Language Models we apply in this work.}
    \vspace{-2ex}
  \label{tab:LLMs_info}%
\end{table*}

\subsubsection{\textbf{Large Language Models As Is}}
We select eight code domain LLMs, i.e., CodeGen25 \cite{codegen2_2023nijkamp}, DeepSeek-Coder \cite{deepseekcoder}, two versions of WizardCoder \cite{wizardcoder2023luo}, four versions of CodeLlama \cite{codellama2023rozière}, and select eight general domain LLMs, i.e., ChatGLM \cite{chatglm2023glm130b}, two versions of Vicuna \cite{vicuna2023judging}, two versions of Llama \cite{Llama2023llama}, two versions of Mistral \cite{mixtral2024jiang}, and ChatGPT \cite{chatgpt2022training}. The principles for our selection are: (1) state-of-the-art, (2) pre-trained on enough source code to be able to understand code to a certain extent, and (3) have text generation and code generation capabilities. In addition, in order to ensure that the model can follow the instructions, we all choose the instruct-tuned version. Table~\ref{tab:LLMs_info} provides detailed information, including parameter size, base model, training corpus, publisher, etc.

Limited by the context window length, we set the maximum input tokens to 4,096, and code snippets exceeding the length will be truncated. For the function name recovery task, we set the maximum new tokens to 48, and for the code summarization task, we set it to 256. We set the sampling temperature to 1, top\_p to 0.95, top\_k to 10, and num\_beams to 1. For all open-source models, we downloaded them from HuggingFace \cite{HuggingFace} and deployed on our local machine with FP16 mixed precision enabled during inference. For ChatGPT, we called its latest \texttt{gpt-3.5-turbo-16k} version through the OpenAI's API.

\subsubsection{\textbf{Prompt Formats}}
Figure \ref{fig:eva_overview} illustrates the prompt format we used for LLMs. We use role-play \cite{chen2023unleashing, kong2024better} prompts to give LLMs the role of experienced binary reverse engineers, enabling them to better handle binary code understanding tasks. Considering the limitation of the length of the model context window, and in order to reduce the inference time overhead and memory usage, we adopt the zero-shot prompts. We analyze the impact of few-shot prompts on the performance of LLMs in Section \ref{subsec:performance_factors}.

\subsubsection{\textbf{Fine-tuned Large Language Models}} \label{subsec:finetunellm}

We also investigate the ability of fine-tuned LLMs to understand binary code, since fine-tuning is a common technique to adapt a pre-trained LLM to downstream tasks \cite{HowEffective_ISSTA_2023,feng2020codebert,fried2023incoder}, such as function name recovery and code summary generation. Furthermore, pre-training corpora of existing LLMs contain very few binary code, either in the form of disassembled instruction sequences or decompiled pseudo code. Therefore, we hope to explore whether injecting binary domain information into LLMs can improve its performance.

The GNU repository\footnote{\url{http://ftp.gnu.org/gnu}} is extensively used as a training or test set for many existing deep learning-based works \cite{nero2020, symlm2022ccs, Chen2023pst, Xiong2023ase,jTrans2022wang, PalmTree21li}. To build our fine-tuning dataset, we select 51 projects from the GNU repository, including binutils, coreutils, findutils, libmicrohttpd, nettle, etc. We use BinKit\cite{Kim2023tse} to create a compilation environment, and then compiler the selected projects using the \textit{gcc-11.2.0} compiler with \textit{x86\_64} target architecture and \textit{O0} optimization level. We obtained a total of 270 binary files. After strip, decompile and alignment, we obtained 124,819 functions matching source code and decompiled pseudo code, and randomly selected 30,000 of them as the fine-tuning dataset.

Additionally, we perform a search and confirm that none of the functions in our proposed benchmark is present in the fine-tuning dataset.

\subsection{Evaluation Setup}
The evaluation environment is a machine equipped with 8 * NVIDIA RTX A6000 GPU with 48GB of VRAM, 2 * 28-core Intel Xeon 6330 CPU, 512GB RAM and 64TB storage, running on Ubuntu 22.04 OS. The GPU is running Nvidia driver version 525.116.03 with CUDA version 12.0. 

We implement all the experiments using Python 3.8 with PyTorch \cite{PyTorch} 2.0.1, DeepSpeed \cite{DeepSpeed} 0.13.0 and Transformers \cite{Transformers} 4.37.2 packages. As for model fine-tuning, we implement it based on the LLaMa-Factory \cite{zheng2024llamafactory} framework. We use LoRA \cite{hu2021lora} fine-tuning method and specify all available modules. We adopt Adam optimizer in fp16 precision, 40 global batch size and 1 training epoch. The learning rate is set to 1e-5 and followed by a consine decay.

\section{Evaluation Results and Findings}

In this section, we conduct extensive experiments to answer the following research questions:

\begin{itemize}
\item \textbf{RQ1:} How do LLMs perform in the task of function name recovery?
\item \textbf{RQ2:} How do LLMs perform in the task of binary code summarization?
\item \textbf{RQ3:} What factors significantly influence the performance of LLMs to understand binary code?
\item \textbf{RQ4:} Can fine-tuning enhance the capability of LLMs to understand binary code?
\item \textbf{RQ5:} Do LLMs have the practical ability in real-world scenarios?
\end{itemize}

\subsection{RQ1: Performance of Function Name Recovery}

\noindent\textbf{Metric.} For the function name recovery task, following \cite{nfre2021issta, Chen2023pst}, we calculate token-level Precision, Recall, F1-score to evaluate the performance of LLMs. The metric ignore non-alphabetical characters and are case-, order-, and duplication-insensitive at the token-level. They can be expressed as:

\begin{table}[t]
  \caption{Comparison of LLMs and DL-based methods on function name recovery. We mark the \colorbox{mycolor}{best} performing methods in each domain.}
  \centering
  \renewcommand{\arraystretch}{1}
  \setlength{\tabcolsep}{0.95mm}{
  \scalebox{0.96}{
    \begin{tabular}{c@{\hspace{0pt}}l@{\hspace{0pt}}cccc}
    \toprule
     \multirow{2}{*}{\textbf{Domain}} & \multicolumn{1}{c}{\multirow{2}{*}{\textbf{Model}}} 
     & \multicolumn{4}{c}{\textbf{Metric}}   \\
     \cmidrule(r){3-6} 
     & & \textbf{\emph{Precision}} & \textbf{\emph{Recall}} & \textbf{\emph{F1-score}} & \textbf{\emph{Time(s)}}  \\
    \midrule
     \multicolumn{1}{c}{\multirow{8}{*}{\makecell{Code \\ LLMs}}} 
       & CodeGen25-7b-instruct & 13.57 & 15.09 & 13.50 & 1.47  \\
       & WizardCoder-15b-V1.0 & 26.41 & 26.14 & 25.40 & 2.22  \\
       & WizardCoder-33b-V1.1 & 21.89 & 22.48 & 21.19 & 5.09  \\
       & Code Llama-7b-instruct-hf & 28.31 & 27.21 & 26.87 & \cellcolor{mycolor}{0.74} \\
       & Code Llama-13b-instruct-hf & 28.26 & 27.30 & 26.76 & 1.60  \\
       & Code Llama-34b-instruct-hf & \cellcolor{mycolor}{28.96} & \cellcolor{mycolor}{28.44} & \cellcolor{mycolor}{27.59} & 3.69  \\
       & Code Llama-70b-instruct-hf & 26.92 & 27.42 & 26.19 & 9.76  \\
       & DeepSeek-Coder-33b-instruct & 25.30 & 27.32 & 25.22 & 4.11 \\
    \midrule
      \multicolumn{1}{c}{\multirow{7}{*}{\makecell{General \\ LLMs}}} 
       & ChatGLM2-6B & 13.98 & 16.36 & 14.12 & \cellcolor{mycolor}{0.59} \\
       & Vicuna-7b-v1.5 & 22.35 & 21.36 & 21.16 & 0.61 \\
       & Vicuna-13b-v1.5 & 25.26 & 24.89 & 24.34 & 0.96 \\
       & Llama-2-13b-chat-hf & 26.78 & \cellcolor{mycolor}{28.31} & \cellcolor{mycolor}{26.69} & 1.19 \\
       & Llama-2-70b-chat-hf & \cellcolor{mycolor}{27.45} & 27.14 & 26.68 & 4.68  \\
       & Mistral-7B-Instruct-v0.2 & 23.38 & 27.49 & 24.28 & 0.84 \\
       & Mixtral-8x7B-Instruct-v0.1 & 19.77 & 25.62 & 21.26 & 2.13  \\
       & ChatGPT(gpt-3.5-turbo-16k) & 22.85 & 22.26 & 21.85 & 1.43 \\
    \midrule
    \multicolumn{1}{c}{\multirow{3}{*}{DL-based}} 
        & SymLM \cite{symlm2022ccs} & 10.69 & 6.04 & 7.72 & 0.29 \\
        & NER \cite{Chen2023pst} & \cellcolor{mycolor}{17.10} & \cellcolor{mycolor}{12.03} & \cellcolor{mycolor}{13.33} & \cellcolor{mycolor}{0.03} \\
        & HexT5 \cite{Xiong2023ase} & 3.69 & 3.33 & 3.28 & 0.17 \\
    \bottomrule
    \end{tabular} } }
    \vspace{0ex}
  \label{tab:funcname_result}%
\end{table}

\begin{equation}
\begin{aligned}
\text{Precision}=\frac{\text{TP}}{\text{TP + FP}}
\end{aligned}
\end{equation}
\begin{equation}
\begin{aligned}
\text{Recall}=\frac{\text{TP}}{\text{TP + FN}}
\end{aligned}
\end{equation}
\begin{equation}
\begin{aligned}
\text{F1-score}=\frac{2 \times \text{Precision} \times \text{Recall}}{\text {Precision}+\text{Recall}}
\end{aligned}
\end{equation}
\noindent\textbf{Result.} Table~\ref{tab:funcname_result} shows the performance of LLMs and DL-based methods on the function name recovery task.

We observe outstanding performance for each parameter size version of CodeLlama and Llama. Among them, CodeLlama-34b outperforms all other LLMs in precision, recall and F1-score metrics, obtaining scores of 28.96, 28.44 and 27.59 respectively. Following closely behind, WizardCoder-15b and DeepSeek-Coder-33b also show excellent performance, obtaining F1-score of 25.40 and 25.22 respectively. CodeGen25-7b and ChatGLM2-6B perform the worst, with only 48.9\% to 66.7\% of other LLMs on the F1-score metric. This lagging performance may be attributed to the capability flaws of their basic models or the lack of targeted training data.

Furthermore, as a task that combines natural language and code language understanding, code domain LLMs generally perform slightly better than general domain LLMs. This may be because their pre-training datasets have a higher proportion of source code. Training on a extensive range of source code datasets allows them to gain a deeper grasp of programming syntax, structure and semantics.

\finding{CodeLLama-34b performs best in the function name recovery task, with an F1-score of 27.59\%. Code domain LLMs generally perform slightly better than general domain LLMs, likely owing to their greater familiarity with programming paradigms.}

Among the deep learning-based expert models, NER \cite{Chen2023pst} performs best, achieving an F1-score of 13.33, but it is still slightly behind CodeGen25, the worst-performing LLM. However, SymLM \cite{symlm2022ccs} and HexT5 \cite{Xiong2023ase} only obtain F1-scores of 7.72 and 3.28, respectively, which differs from the performance reported in their original articles. This difference may come from the partitioning of their datasets. SymLM and HexT5 widely use projects from GNU as part of their training and testing sets. SymLM divides the training and testing sets at the binary file-level, which may result in some code appearing in both the training and testing sets. For example, in the \texttt{Binutils} project, the \texttt{ar} and \texttt{nm} files share the same binary file descriptor(BFD) processing code. This reuse of libraries and underlying code may lead to exaggeration of evaluation metrics. Although HexT5 adopts a stricter project-level dataset partitioning approach, different projects under GNU may still share programming styles or naming conventions, leading to potential data leaks. In summary, these models have limited generalization ability for data outside the training data distribution, and this zero-shot learning ability is exactly what LLMs are good at.

\finding{The existing DL-based expert models have poor generalization ability for data outside the training distribution, and their performance is far lower than the LLMs.}

In terms of inference time cost, locally deployed LLMs with 6B-7B parameter quantities typically require 0.6 to 0.8 seconds to infer a single piece of data, LLMs with 13-15B scales require 1.2 to 2.2 seconds, LLMs with 33-34B scales require 4.1 to 5.1 seconds, and CodeLlama-70b requires a maximum of 9.76 seconds per piece. For ChatGPT that requires API calls, inference for each data takes 1.43 seconds. The DL-based model, due to its lightweight advantage, greatly reduces inference time and achieves the fastest NER of 0.03 seconds per piece.

\finding{The DL-based model has a significant advantage in inference speed benefiting from their model size. Meanwhile, the inference speed of LLMs is still within an acceptable range.}

\subsection{RQ2: Performance of Binary Code Summarization}
\noindent\textbf{Metric.} For the binary code summarization task, same as BinT5 \cite{bint52023saner}, HexT5 \cite{Xiong2023ase}, we use smoothed BLEU-4 \cite{papineni-etal-2002-bleu}, METEOR \cite{meteor2009Lavie}, Rouge-L \cite{lin-2004-rouge} as the evaluation metric.

\noindent\textbf{Result.} The performance of LLMs and DL-based methods on the binary code summarization task is presented in Table~\ref{tab:summarization_result}.

ChatGPT outperforms all other LLMs in BLEU-4, METEOR and Rouge-L metrics, obtaining scores of 7.37, 28.13 and 23.80 respectively. WizardCoder-15b also shows very competitive results, achieving 6.60, 26.53, and 23.80, respectively. Similar to the function name recovery task, CodeGen25-7b and ChatGLM2-6B performed the worst in their respective domains, but narrowed the performance gap with other LLMs.

\begin{table}[t]
  \caption{Comparison of LLMs and DL-based methods on binary code summarization. We mark the \colorbox{mycolor}{best} performing methods in each domain.}
  \centering
  \renewcommand{\arraystretch}{1}
  \setlength{\tabcolsep}{0.92mm}{
  \scalebox{0.95}{
    \begin{tabular}{c@{\hspace{0pt}}l@{\hspace{0pt}}cccc}
    \toprule
     \multirow{2}{*}{\textbf{Domain}} & \multicolumn{1}{c}{\multirow{2}{*}{\textbf{Model}}} & \multicolumn{4}{c}{\textbf{Metric}} 
     \\
     \cmidrule(r){3-6}
     & & \textbf{\emph{BLEU-4}} & \textbf{\emph{METEOR}} & \textbf{\emph{Rouge-L}} & \textbf{\emph{Time(s)}} \\
    \midrule
     \multicolumn{1}{c}{\multirow{8}{*}{\makecell{Code \\ LLMs}}} 
       & CodeGen25-7b-instruct   & 3.87 & 23.61 & 18.95 & 7.32 \\
       & WizardCoder-15b-V1.0  & \cellcolor{mycolor}{6.60} & \cellcolor{mycolor}{26.53} & \cellcolor{mycolor}{23.80} & 12.91 \\
       & WizardCoder-33b-V1.1 & 4.74 & 25.43 & 20.34 & 23.71 \\
       & Code Llama-7b-instruct-hf & 4.65 & 22.20 & 20.66 & \cellcolor{mycolor}{5.33} \\
       & Code Llama-13b-instruct-hf & 4.03 & 19.43 & 19.68 & 8.37 \\
       & Code Llama-34b-instruct-hf & 4.69 & 21.13 & 21.14 & 18.93 \\
       & Code Llama-70b-instruct-hf & 4.64 & 23.75 & 20.82 & 53.55 \\
       & DeepSeek-Coder-33b-instruct & 5.22 & 24.10 & 20.84 & 20.26 \\
    \midrule
      \multicolumn{1}{c}{\multirow{7}{*}{\makecell{General \\ LLMs}}} 
       & ChatGLM2-6B & 4.29 & 26.37 & 20.66 & 5.53 \\
       & Vicuna-7b-v1.5 & 5.99 & 25.29 & 22.84 & 5.54 \\
       & Vicuna-13b-v1.5 & 4.90 & 21.20 & 22.48 & 7.53 \\
       & Llama-2-13b-chat-hf & 6.16 & 24.47 & 22.51 & 11.30 \\
       & Llama-2-70b-chat-hf & 5.51 & 26.26 & 21.51 & 47.76 \\
       & Mistral-7B-Instruct-v0.2 & 5.98 & 24.37 & 23.55 & 3.64 \\
       & Mixtral-8x7B-Instruct-v0.1 & 5.79 & 24.16 & 23.56 & 11.00 \\
       & ChatGPT(gpt-3.5-turbo-16k)  & \cellcolor{mycolor}{7.37} & \cellcolor{mycolor}{28.13} & \cellcolor{mycolor}{23.76} & \cellcolor{mycolor}{2.85}  \\ 
    \midrule
    \multicolumn{1}{c}{\multirow{2}{*}{DL-based}}
        & BinT5 \cite{bint52023saner} & 0.00 & 2.12 & 4.83 & 0.59  \\
        & HexT5 \cite{Xiong2023ase} & \cellcolor{mycolor}{0.09}  & \cellcolor{mycolor}{6.32} & \cellcolor{mycolor}{8.53} & \cellcolor{mycolor}{0.47}  \\
    \bottomrule
    \end{tabular} }
    \vspace{0ex}
  \label{tab:summarization_result} }
\end{table}

Unlike the function name recovery task, the performance of general domain LLMs is generally significantly better than that of code domain LLMs in binary code summary tasks. This may be attributed to the different properties of the two tasks. In the function name recovery task, the output of LLMs is usually shorter and only needs to generate a function name, which is relatively simple. In contrast, the binary code summarization task requires generating longer natural language descriptions to accurately summarize the functionality and structure of binary code, which requires the model to understand more contextual information and convert it into natural language text, which is a more complex task. General domain LLMs are better at generating longer natural language descriptions due to their inherent characteristics, while code domian LLMs have limited capabilities in this regard.

\finding{ChatGPT performs best among all LLMs in the binary code summarization task with a BLEU-4 of 7.37\%. General domain LLMs perform significantly better than code domain LLMs, which is attributed to its stronger long-context understanding and summarizing capabilities.}

For the DL-based expert model, BinT5 \cite{bint52023saner} obtains 0.00, 2.12 and 4.83 on the BLEU-4, METEOR, Rouge-L metric, and HexT5 \cite{Xiong2023ase} improves slightly, obtaining 0.09, 6.32 and 8.53 respectively, but the performance is still far lower than LLMs.

\finding{Similar to the previous task, existing DL-based expert models perform worse than LLMs on the binary code summarization task.}

Regarding inference time, locally deployed LLMs are generally 5-6 times longer than the function name recovery task. LLMs with 6B-7B parameters usually take 3.6 to 7.3 seconds to infer a single piece of data, 13-15B scale LLMs take 7.5 to 12.9 seconds, and 33-34B scale LLMs takes 18.9 to 23.7 seconds. CodeLlama-70b takes the most of 53.55 seconds among LLMs. For ChatGPT, which requires API calls, inference takes 2.85 seconds per data, roughly twice as long as the function name recovery task. The DL-based model also shows the advantage of inference speed, with HexT5 taking only the shortest 0.47 seconds.

\finding{For the binary code summarization task, inference time for locally deployed LLMs is about five times that of function name recovery.}
\vspace{-0.5ex}

\subsection{RQ3: Factors that Significantly Affect Performance}\label{subsec:performance_factors}

\begin{table*}[t]
  \caption{Performance of prompts in the form of Few-shot. The Impr. column represents the performance improvement of Few-shot compared to Zero-shot. We mark the \colorbox{mycolor}{increase} and \colorbox{gray!30}{decrease} of the metrics.}
  \centering
  \renewcommand{\arraystretch}{1.15}
  \setlength{\tabcolsep}{1mm}{
  \scalebox{0.95}{
    \begin{tabular}{l@{\hspace{3pt}}cccccccccccccccc}
    \toprule
     \multicolumn{1}{c}{\multirow{4}{*}{\textbf{Model}}} & \multicolumn{8}{c}{\textbf{Function Name Recovery}} & \multicolumn{8}{c}{\textbf{Binary Code Summarization}} 
     \\
     \cmidrule(r){2-9}  \cmidrule(r){10-17}
     & \multicolumn{2}{c}{\textbf{\emph{Precision}}} &  \multicolumn{2}{c}{\textbf{\emph{Recall}}} &  \multicolumn{2}{c}{\textbf{\emph{F1-score}}} &  \multicolumn{2}{c}{\textbf{\emph{Time(s)}}} &  \multicolumn{2}{c}{\textbf{\emph{BLEU-4}}} &  \multicolumn{2}{c}{\textbf{\emph{METEOR}}} &  \multicolumn{2}{c}{\textbf{\emph{Rouge-L}}} &  \multicolumn{2}{c}{\textbf{\emph{Time(s)}}} \\
     \cmidrule(r){2-3}  \cmidrule(r){4-5} \cmidrule(r){6-7}  \cmidrule(r){8-9} \cmidrule(r){10-11}  \cmidrule(r){12-13} \cmidrule(r){14-15}  \cmidrule(r){16-17}
    & \textbf{Few} & \textbf{Impr.}  & \textbf{Few} & \textbf{Impr.} & \textbf{Few} & \textbf{Impr.} & \textbf{Few} & \textbf{Impr.} & \textbf{Few} & \textbf{Impr.} & \textbf{Few} & \textbf{Impr.} & \textbf{Few} & \textbf{Impr.} & \textbf{Few} & \textbf{Impr.} \\
    \midrule
        WizardCoder-15b-V1.0 & 30.65 & \cellcolor{mycolor}{+4.24pt} & 30.90 & \cellcolor{mycolor}{+4.76pt} & 29.17 & \cellcolor{mycolor}{+3.77pt} & 1.82 & \cellcolor{gray!30}{-0.40s} & 7.30 & \cellcolor{mycolor}{+0.70pt} & 29.69 &\cellcolor{mycolor}{+3.16pt} & 24.52 & \cellcolor{mycolor}{+0.72pt} & 11.32 & \cellcolor{gray!30}{-1.59s}  \\
        
        Code Llama-7b-instruct-hf & 29.64 & \cellcolor{mycolor}{+1.33pt} & 27.86 & \cellcolor{mycolor}{+0.65pt} & 28.01 & \cellcolor{mycolor}{+1.14pt} & 1.13 & \cellcolor{mycolor}{+0.39s} & 6.62 & \cellcolor{mycolor}{+1.97pt} & 27.64 &\cellcolor{mycolor}{+5.44pt} & 23.58 & \cellcolor{mycolor}{+2.92pt} & 6.69 & \cellcolor{mycolor}{+1.36s}  \\
    \midrule
        Vicuna-7b-v1.5 & 27.24 & \cellcolor{mycolor}{+4.89pt} & 25.44 & \cellcolor{mycolor}{+4.08pt} & 25.52 & \cellcolor{mycolor}{+4.36pt} & 0.88 & \cellcolor{mycolor}{+0.17s} & 6.68 & \cellcolor{mycolor}{+0.69pt} & 22.39 &\cellcolor{gray!30}{-2.90pt} & 21.35 & \cellcolor{gray!30}{-1.49pt} & 4.43 & \cellcolor{gray!30}{-1.11s}  \\
        
        Mistral-7B-Instruct-v0.2 & 31.45 & \cellcolor{mycolor}{+8.07pt} & 31.87 & \cellcolor{mycolor}{+4.38pt} & 30.73 & \cellcolor{mycolor}{+6.45pt} & 1.06 & \cellcolor{mycolor}{+0.22s} & 7.94 & \cellcolor{mycolor}{+1.96pt} & 29.32 &\cellcolor{mycolor}{+4.95pt} & 25.00 & \cellcolor{mycolor}{+1.45pt} & 5.96 & \cellcolor{mycolor}{+2.32s}  \\
        
        ChatGPT(gpt-3.5-turbo-16k) & 32.96 & \cellcolor{mycolor}{+10.11pt} & 31.17 & \cellcolor{mycolor}{+8.91pt} & 31.24 & \cellcolor{mycolor}{+9.39pt} & 1.79 & \cellcolor{mycolor}{+0.36s} & 9.71 & \cellcolor{mycolor}{+2.34pt} & 32.13 & \cellcolor{mycolor}{+4.00pt} & 27.21 & \cellcolor{mycolor}{+3.45pt} & 3.50 & \cellcolor{mycolor}{+0.65s}  \\
    \bottomrule
    \end{tabular} } 
    \vspace{0ex}
  \label{tab:fewshot_result} }
\end{table*} 

We further explore the key factors affecting the performance of LLMs, including the few-shot form prompts, the length of pseudo code, and the length of symbol information.

\subsubsection{\textbf{Few-shot prompts}}
The pre-training datasets for LLMs contain little or no binary code, which makes directly applying LLMs to binary code understanding tasks likely not to yield optimal results. In this case, few-shot prompts become a potential solution, by providing well-designed examples to LLMs, so that LLMs can learn the unique structure and syntax of binary code and quickly adapt to new tasks. Specifically, we construct two carefully designed pairs of pseudo code and ground-truth examples, and add them to the original prompts. We select five LLMs that performed well in the previous experiments to conduct the few-shot prompts experiment. The results are shown in Table~\ref{tab:fewshot_result}. 

For the function name recovery task, compared with the zero-shot prompts, general domain LLMs have a more obvious improvement compared to code domain LLMs. Among them, ChatGPT has improved by 10.11, 8.91, and 9.39 points in Precision, Recall and F1-score metric respectively. For the binary code summarization task, the LLMs of both domains have improved slightly, among which ChatGPT has improved by 2.34, 4.00, and 3.45 points in the BLEU-4, METEOR, and Rouge-L metric respectively. The exception is that Vicuna-7b shows a slight decrease in the METEOR and Rough-L metrics. Observing its output, we found that the introduction of the few-shot examples increased the length of prompts, causing more test data to exceed the maximum length of window context of the model (4096 tokens) and be truncated, resulting in a decrease in performance.

In addition, few-shot prompts will improve inference time in most cases. However, observing the outputs of WizardCoder-15b, we found that the few-shot prompts improve the model's ability to follow instructions, reduce the output of useless information, and thus reduce the inference time. In this case, few-shot prompts not only enhance model performance but also improve inference efficiency.

\finding{When computing resources and inference time permit, few-shot prompts can be selected to improve the performance of LLMs on function name recovery and binary code summarization tasks.}

\begin{figure}
    \centering
    \begin{subfigure}[b]{0.49\linewidth}
        \centering
        \includegraphics[width=\linewidth]{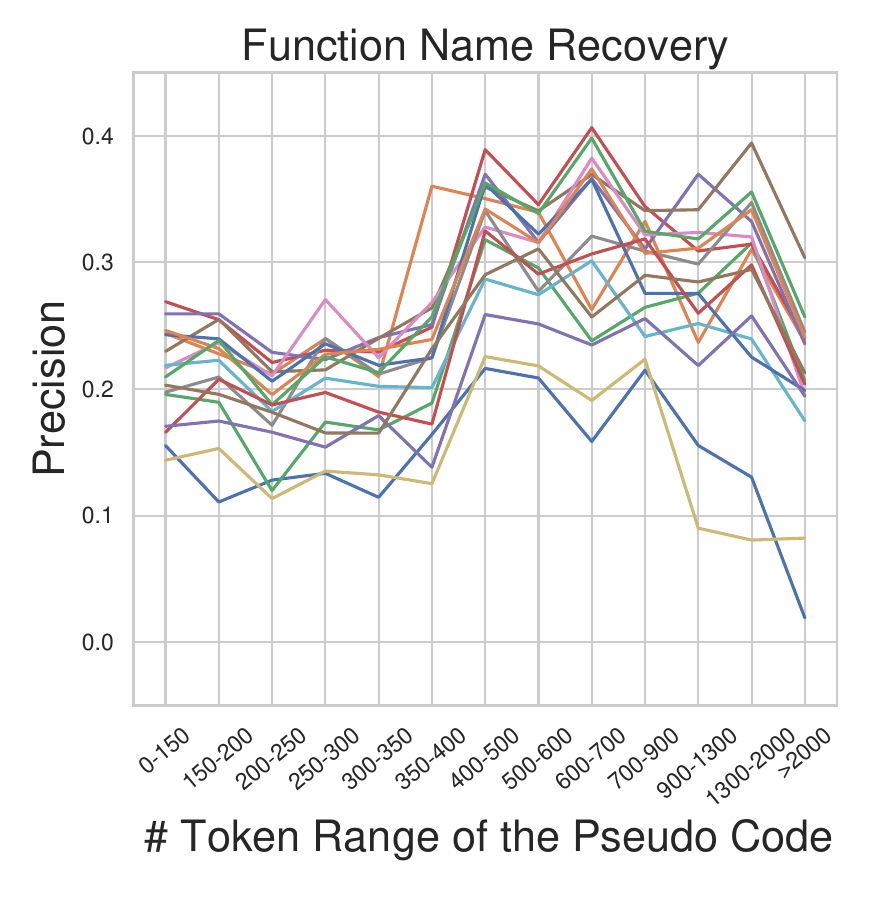}   
    \end{subfigure}
    \begin{subfigure}[b]{0.49\linewidth}
        \centering
        \includegraphics[width=\linewidth]{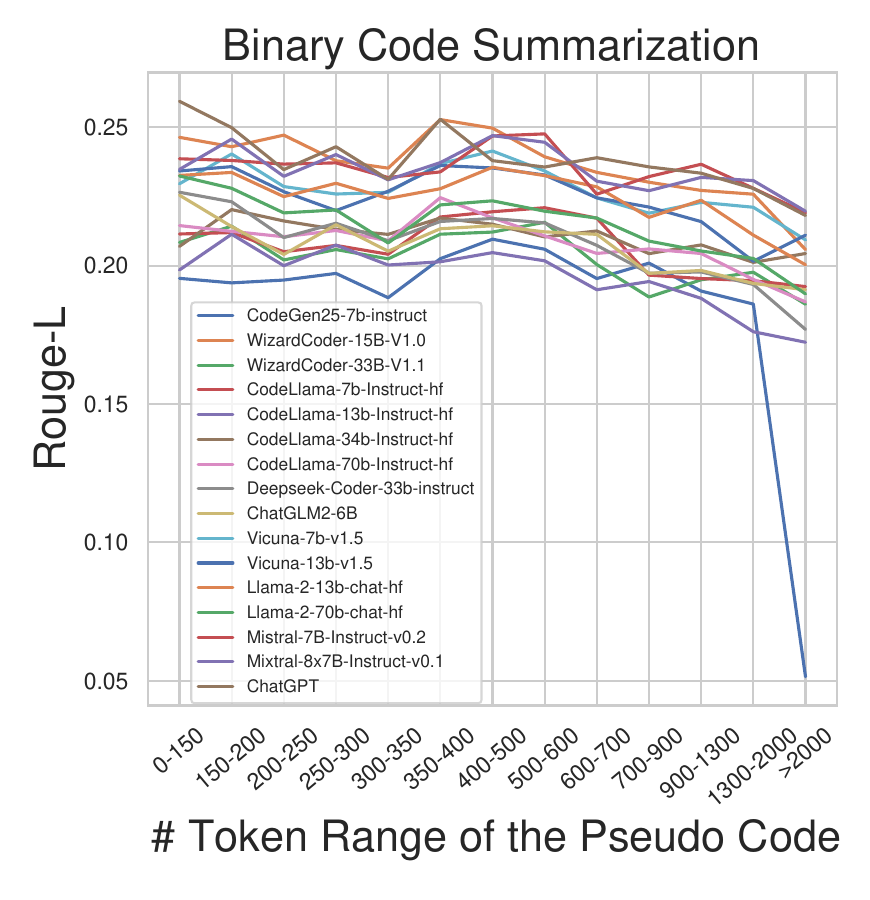}
    \end{subfigure}
    \caption{Impact of pseudo code length on performance.}
    \vspace{-1.2ex}
    \label{fig:pcodelen}
\end{figure}

\subsubsection{\textbf{Pseudo Code Length}}\label{subsec:pcodelensec}
To study the impact of pseudo code length on the performance of LLMs, we divide the length of pseudo code token according to intervals, and controll the number of data in each interval between 100 and 200 to avoid long-tail distribution of data.

As shown in Figure \ref{fig:pcodelen}, when the length of pseudo code is between 0-400 tokens, the metric of function name recovery remains at a relatively low level, as shorter pseudo code may not provide enough keywords to infer the purpose and naming intention of the function. Longer pseudo code can provide more contextual information, helping the LLMs capture semantic clues related to function names. Therefore, the metrics are relatively high between 400-2000 tokens; After exceeding 2000 tokens, the structure and logic of the code are too complex, making it difficult for the LLMs to process and integrate a large amount of information, resulting in a decrease in metrics.

For the binary code summarization task, the metrics show a slowly decreasing trend as the pseudo code length increases. As code complexity increases, LLMs find it difficult to maintain both conciseness and accuracy of summaries resulting in the generation of lengthy and unfocused summaries, thereby reducing the overall quality of the summaries.

\finding{LLMs achieve the best performance for function name recovery at moderate pseudo code length, while the performance of binary code summarization slowly decreases as pseudo code length increases.}

\begin{figure}
    \centering
    \begin{subfigure}[b]{0.49\linewidth}
        \centering
        \includegraphics[width=\linewidth]{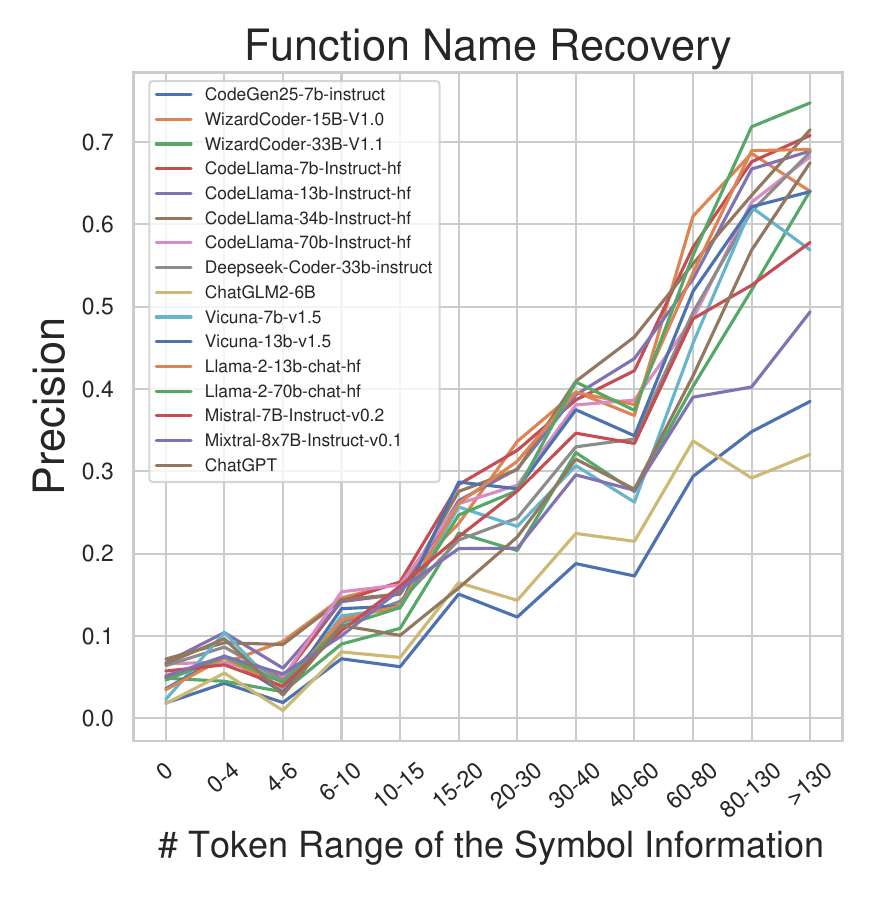}   
    \end{subfigure}
    \begin{subfigure}[b]{0.49\linewidth}
        \centering
        \includegraphics[width=\linewidth]{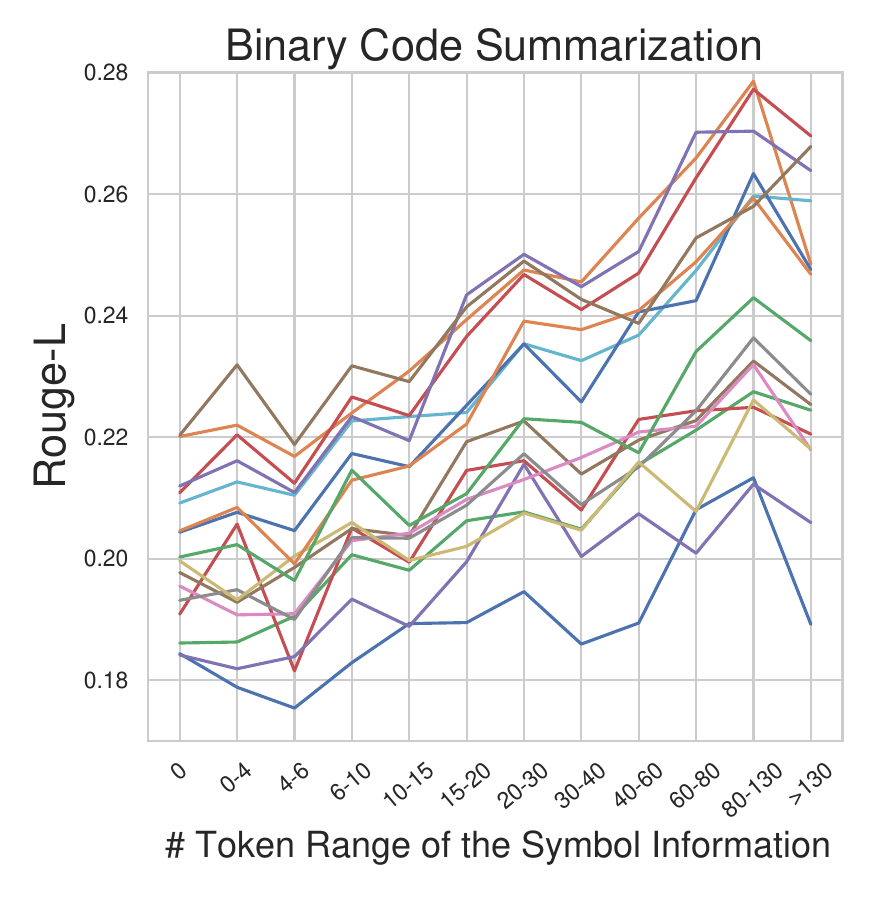}
    \end{subfigure}
    \caption{Impact of symbol information length on performance.}
    \vspace{-2ex}
    \label{fig:symbollen}
\end{figure}

\subsubsection{\textbf{Symbol Information Length}}
We define symbol information as the strings and identifiers in the pseudo code that are not stripped during the strip process, which can provide human-understandable semantic information. We also divide the length of the symbolic information token into intervals.

As shown in Figure \ref{fig:symbollen}, as the length of the symbol information token increases, the performance of both function name recovery and binary code summarization tasks increases significantly. This is due to the fact that longer symbol information provides richer semantic content and more context clues, helping LLMs understand the intent and functionality of the code. However, we found that when the symbol information token exceeds 130, the performance of most LLMs in binary code summarization tasks slightly decreases. This is because more symbol information tokens are accompanied by longer pseudo code lengths, resulting in more code being truncated due to window context limitations, affecting the completeness of LLMs summary.

\finding{The symbol information (e.g., strings and identifiers) has rich semantics and contributes significantly to LLM's understanding of binary code.} 
\vspace{-0.5ex}

\subsection{RQ4: Fine-Tuning to Enhance the Performance}

As mentioned in Section \ref{subsec:finetunellm}, we build the fine-tuning dataset from the GNU repository, with each piece of data in the form of decompiled pseudo code and ground-truth pairs. Considering the computational resource limitations, we only fine-tune the 7b LLMs, and we choose \texttt{Codellama-7b-instruct-hf} of the code domain and \texttt{Mistral-7B-Instruct-v0.2} of the general domain that have performed well in previous experiments. 

\begin{figure}
    \centering
    \begin{subfigure}[b]{0.9\linewidth}
        \centering
        \includegraphics[width=\linewidth]{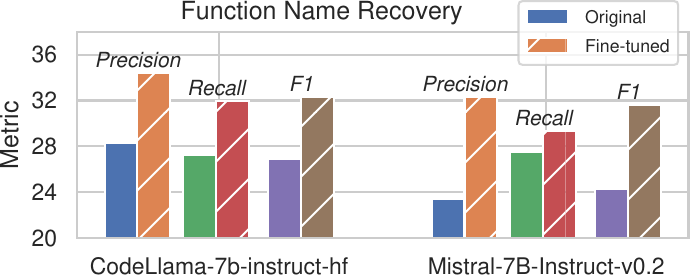}
    \end{subfigure}
    \vspace{3pt}
    \begin{subfigure}[b]{0.9\linewidth}
        \centering
        \includegraphics[width=\linewidth]{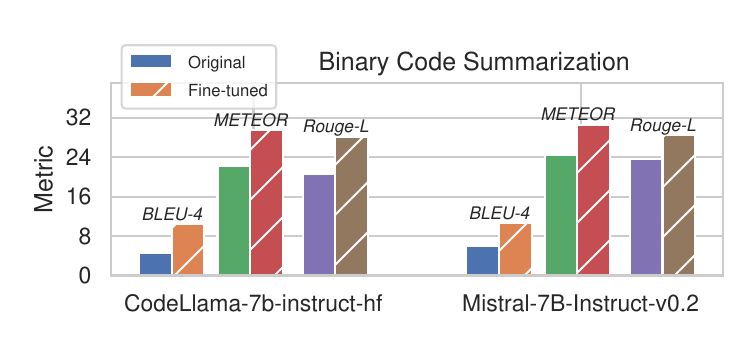}
    \end{subfigure}
    \caption{Comparison of original and fine-tuned LLMs on performance.}
    \vspace{-2ex}
    \label{fig:finetune}
\end{figure}

Figure \ref{fig:finetune} shows the performance comparison of original and fine-tuned LLMs. For the function name recovery task, the Precision, Recall and F1-score metrics increase by an average of 7.50, 3.32 and 6.37 points respectively. For the binary code summarization task, the BLEU-4, METEOR and Rouge-L metrics increase by an average of 5.20, 6.74 and 6.18 points respectively. Overall, fine-tuning LLMs on downstream tasks related to binary code understanding can bring considerable performance improvements.

\finding{Introducing binary domain knowledge through fine-tuning can improve the performance of LLMs on function name recovery and summary production tasks.}
\vspace{-1.5ex}

\subsection{RQ5: Case Study on Real-World Virus Analysis \label{sec:case_study}}

\begin{figure*}
    \centering
    \begin{subfigure}[b]{0.85\linewidth}
        \includegraphics[width=1\linewidth]{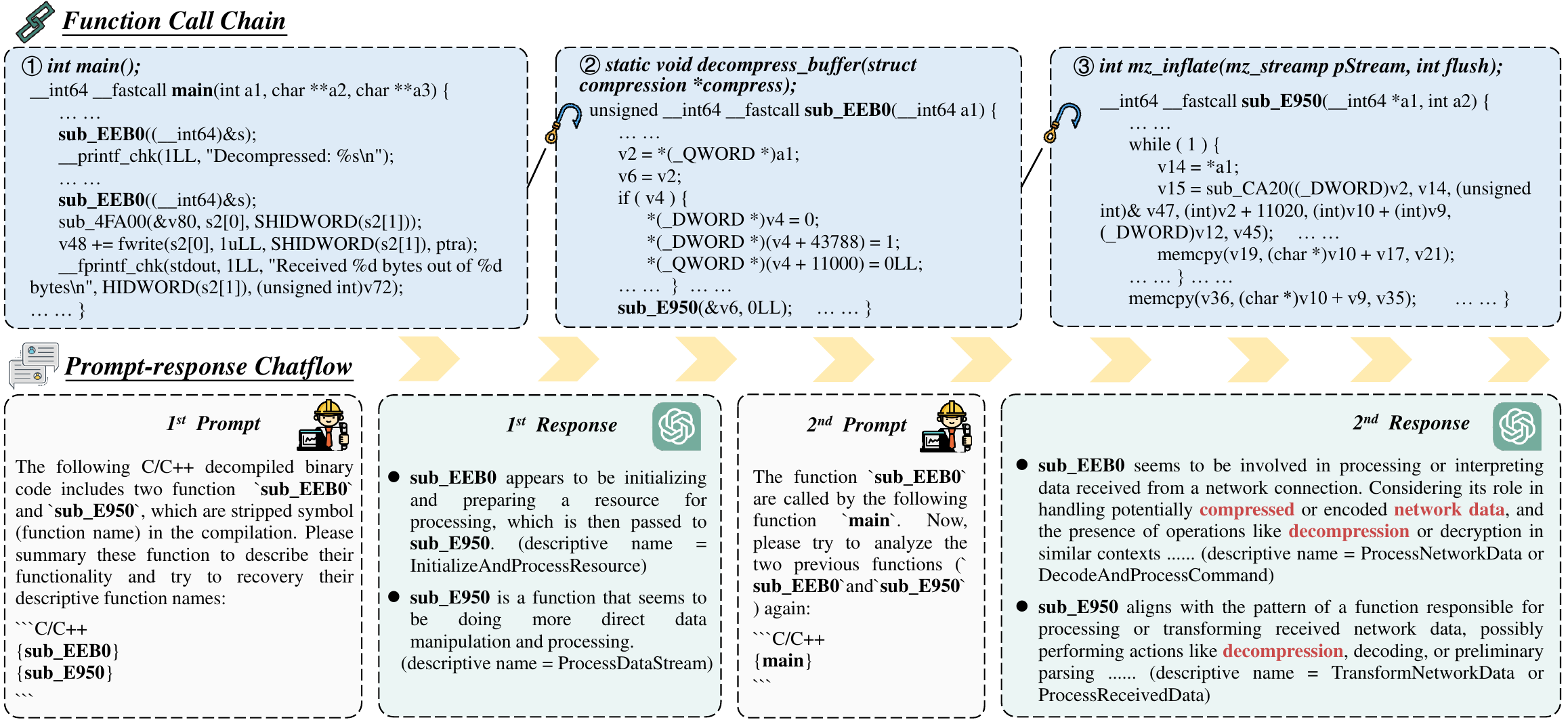}
    \end{subfigure}
    \vspace{0ex}
    \caption{An example of binary code understanding in a real-world virus with ChatGPT.}
    \vspace{-2ex}
    \label{fig:case_splinter}
\end{figure*}

We present a case study to show how much advanced general LLMs can assist participants in a real-world scenario. Specifically, we utilize ChatGPT \cite{chatgpt2022training} to facilitate virus analysis, including summarizing the functionality of decompiled binary functions in viruses and recovering their descriptive names. 

An open-source Linux remote access trojan named \texttt{splinter}\footnote{\url{https://github.com/tuian/splinter}} is compiled with \textit{gcc-11.4.0} and stripped to release. In this case, Figure \ref{fig:case_splinter} has shown a partial analysis in a call chain, where the reduced pseudo code are shown in the upper part, as well as the function definitions in the source code. The lack of symbolic information makes it difficult for an analyst to understand the function \texttt{sub\_EEB0} and \texttt{sub\_E950}. We first constructed the first prompt with these two functions and fed it to ChatGPT, asking for the functional summaries and descriptive names. The model gives us a primary description of the operations performed in the pseudo code without any high-level insight. We then constructed the second prompt with the caller function \texttt{main} that contains a few symbol information. The second response correctly indicates that these functions are related to data decompression. At the same time, the predicted names reflect their functionality, although the predictions do not exactly match the source code. 

\finding{ChatGPT demonstrates the potential ability to analyze binaries in the real world. The information from the calling context will boost the predictions of LLMs.} 
\vspace{-1.2ex}

\section{Discussions}

Based on the experimental findings, we summarize directions for future work and limitations of the current evaluation.
\vspace{-1.2ex}
\subsection{Future Works}
Current LLMs have indeed shown potential in understanding binary codes, and we believe that future work can be conducted in-depth from the following aspects.

\begin{itemize}	
	\item \emph{Develop domain-specific LLM}: craft and refine LLMs tailored for binary code analysis and understanding, incorporating extensive binary domain knowledge during the pre-training phase to enhance the LLMs's grasp of code semantics and structure.
	
	\item \emph{Extend context window}: investigate advanced architectures and techniques that support longer sequence lengths, allowing LLMs to effectively analyze and comprehend extensive code snippets. 
	
	\item \emph{Enhance processing of non-intuitive code}: address the current shortfall of LLMs in handling code devoid of descriptive strings or identifiers by devising innovative algorithms that can decipher and interpret the functionality of complex, non-obvious code snippets.

    \item \emph{Integrate Multi-Modal Information}: explore strategies for integrating diverse information sources, such as expert human annotations, assembly instructions, and dynamic execution data, into the LLMs' input to provide a comprehensive understanding of binary code.
\end{itemize}

\label{sec:bibtex}

\vspace{-1.2ex}
\subsection{Limitations}
Although this paper provides a systematic evaluation of the performance of LLMs on binary code understanding tasks, we need to acknowledge existing limitations.
\begin{itemize}	
	\item \emph{Evaluation metrics for code summarization tasks:} Current practices predominantly utilize text coherence metrics like BLEU-4 \cite{papineni-etal-2002-bleu} and Rouge-L \cite{lin-2004-rouge}, which are originally designed for text translation tasks. However, these may not be entirely appropriate for binary code summarization. Reverse engineers can usually understand the specific design of a function through certain key terms, and text fluency is not that important at this time. It may be beneficial to develop a new metric to better capture the essence of binary summaries.
	
	\item \emph{Binary code obfuscation:} Our evaluation dataset does not consider any form of binary code obfuscation, such as encryption or compiler-based obfuscation \cite{Junod2015obf}. In fact, obfuscation can significantly change the form of pseudo code, making it challenging for models to accurately understand and interpret the code. This may affect the applicability of LLMs in real-world scenarios where confusion is prevalent.
\end{itemize}

\label{sec:bibtex}

\vspace{-1ex}
\section{Conclusion}
In this paper, we selected two representative tasks: (1) function name recovery, and (2) binary code summarization, and designed an automated method to construct a benchmark to comprehensively evaluate the capabilities of Large Language Models (LLMs) to understand binary code. 

Our findings indicate that LLMs demonstrate promising capabilities in advancing automated binary code understaning, and we call for more new research to focus on this important domain of software engineering to further enhance the capabilities of LLMs to play a more critical role in the complex task of binary code analysis.

\section*{Acknowledgement}
We thank the reviewers for their insightful comments and suggestions. This work was supported in part by the Natural Science Foundation of China under Grant U20B2047, 62072421, 62002334, 62102386 and 62121002, and by Open Fund of Anhui Province Key Laboratory of Cyberspace Security Situation Awareness and Evaluation under Grant CSSAE-2021-007.

\bibliographystyle{IEEEtran}
\bibliography{ref}

\end{document}